\documentclass[12pt,preprint]{aastex}

\def\deg{\hbox{$^{\:\circ}$}}

\slugcomment{}

\shorttitle{Locating the Silicate Emission in NGC~2110}
\shortauthors{Mason et al.}

\begin{document}

\title{The Origin of the Silicate Emission Features in the Seyfert 2 Galaxy, NGC2110}

%% Use \author, \affil, and the \and command to format
%% author and affiliation information.
%% Note that \email has replaced the old \authoremail command
%% from AASTeX v4.0. You can use \email to mark an email address
%% anywhere in the paper, not just in the front matter.
%% As in the title, you can use \\ to force line breaks.

\author{R. E. Mason}
\affil{Gemini Observatory, Northern Operations Center, 670 N. A'ohoku Place, Hilo, HI 96720}
\email{rmason@gemini.edu}

\author{N. A. Levenson}
\affil{Department of Physics and Astronomy, University of Kentucky, Lexington, KY 40506}

\author{Y. Shi}
\affil{Steward Observatory, University of Arizona, 933 North Cherry Avenue, Tucson, AZ 85721}

\author{C. Packham}
\affil{University of Florida, Department of Astronomy, 211 Bryant Space Science Center, Gainesville, FL 32611}

\author{V. Gorjian}
\affil{Jet Propulsion Laboratory, California Institute of Technology, 4800 Oak Grove Drive, Pasadena, CA 91109}

\author{K. Cleary}
\affil{Jet Propulsion Laboratory, California Institute of Technology, 4800 Oak Grove Drive, Pasadena, CA 91109}

\author{J. Rhee}
\affil{Department of Physics and Astronomy, University of California, Los Angeles, CA 90095}

\and

\author{M. Werner}
\affil{Jet Propulsion Laboratory, California Institute of Technology, 4800 Oak Grove Drive, Pasadena, CA 91109}

\begin{abstract}

The unified model of active galactic nuclei (AGN) predicts silicate emission features at 10 and 18~$\mu$m in type 1 AGN, and such features have now been observed in objects ranging from distant QSOs to nearby LINERs. More surprising, however, is the detection of silicate emission in a few {\it  type 2} AGN. By combining Gemini and {\it Spitzer} mid-infrared imaging and spectroscopy of NGC~2110, the closest known Seyfert 2 galaxy with silicate emission features, we can constrain the location of the silicate emitting region to within 32 pc of the nucleus. This is the strongest constraint yet on the size of the silicate emitting region in a Seyfert galaxy of any type. While this result is consistent with a narrow line region origin for the emission, comparison with clumpy torus models demonstrates that emission from an edge-on torus can also explain the silicate emission features and 2--20~$\mu$m spectral energy distribution of this object.  In many of the best-fitting models the torus has only a small number of clouds along the line of sight, and does not extend far above the equatorial plane. Extended silicate-emitting regions may well be present in AGN, but this work establishes that emission from the torus itself is also a viable option for the origin of silicate emission features in active galaxies of both type 1 and  type 2.

\end{abstract}


\section{Introduction}

In the unified model of active galactic nuclei (AGN), the differences between objects of types 1 and 2 are explained by a torus of dust and gas that obscures the broad line gas from some viewing angles while leaving it exposed from others. Various lines of evidence, notably the detection of polarized broad emission lines in type 2 objects  \citep{AM85}, support the hypothesis of anisotropic obscuration, but the chemistry, structure and origin of the torus, as well as the true extent of its role in unifying the various types of AGN, remain the subject of intense study.

The dust in the torus is expected to absorb short wavelength radiation from the 
nucleus and re-emit it in the mid-infrared (MIR), and a strong constraint on models aiming to explain 
and predict emission from the torus is the Si-O bond stretch near 10~$\mu$m. In the simplest realisations of the unified model, an edge-on view through cool dust in the torus causes a prominent absorption feature in type 2 AGN. In type 1 AGN, where hot dust at the inner surface of the torus is revealed, models often predict strong silicate emission \citep[e.g.][]{Efstathiou95,Granato94}, depending on the torus geometry \citep{PK92}.

In practice, however, the situation has proved less straightforward than this. For instance, in a sample of 16 Seyfert 1 galaxies, \citet{Roche91} detected silicate emission in only one. More recently, Spitzer observations of QSOs have established not only that the silicate emission feature is common in those objects \citep{Hao05,Hao07,Siebenmorgen05} but also  - surprisingly - that silicate 
emission is present in the average type 2 QSO spectrum \citep{Sturm06}. \citet{Shi06} find that in a Spitzer/IRS sample of $\sim$100 AGN, X-ray column density roughly correlates with 
silicate feature strength, in agreement with unified schemes. However, there are some obvious anomalies: type 1 AGN with strong silicate absorption and type 2s with silicate emission. These observations are not readily explained by simple torus models.

\begin{figure}[ht]
\includegraphics[scale=0.65,angle=270]{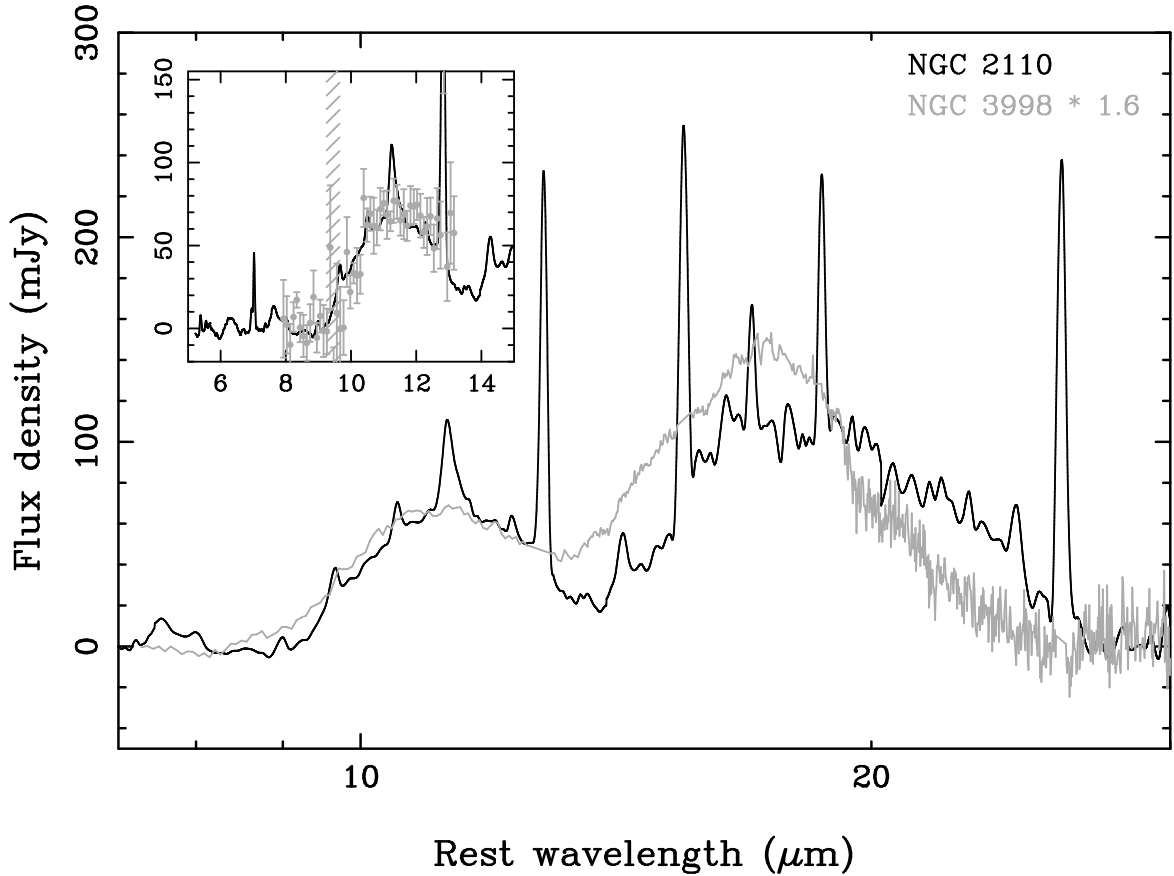}
\caption{Continuum-subtracted Michelle and IRS spectra of NGC~2110 (inset), and continuum-subtracted IRS spectra of NGC~2110 and NGC~3998 \citep[][emission lines removed]{Sturm05}. The NGC~3998 spectrum has been multiplied by 1.6 and the data in the main figure displayed on a logarithmic wavelength scale to emphasize the feature profiles. }
\label{fig:specs2}
\end{figure}

\citet{Sturm05} note that the dust temperature implied by the silicate bands in the QSO spectra is $\sim$200~K, suggestive of dust in the narrow-
line region (NLR) rather than in the hot inner wall of the torus. NLR dust with a temperature of $\sim$200-300~K has been detected in mid-IR images of several AGN, extended over $\sim$1-2\arcsec\ from the unresolved nucleus \citep[e.g.][]{Bock2,Radomski03,Gorjian04,Packham05}. Such dust, heated by the central engine but not directly associated with the torus itself, could explain the unexpected silicate emission in type 2 objects. Alternatively, alterations to the basic torus models may enable them to match the observations. Several recent torus models have incorporated a clumpy dust distribution {\citep{Nenkova,Dullemond05,Honig06,Schartmann08}, which is expected from considerations of grain survival in the circumnuclear environment \citep{Krolik88}  and whose existence is suggested by interferometric observations of the Circinus galaxy \citep{Tristram07}. In a clumpy torus, cool dust may be present close to the central engine.
\citet{Nenkova08b} find that in certain circumstances a clumpy torus can produce silicate emission features even for lines of sight close to the equatorial plane, especially if the optical depth per cloud $\tau_{V}>100$ or if the number of clouds along the line of sight is as low as $N_{0}\sim 2$. The silicate emission in type 2 AGN could therefore come from either the NLR or the torus.

 While Spitzer's excellent sensitivity is well suited to discovering silicate emission in AGN, its spatial resolution 
 cannot pinpoint the location of the emitting dust. Ground-based mid-IR spectra of NGC~1068 show that spatial resolution is critical: the depth and profile of the silicate feature vary considerably on sub-arcsecond ($<$70 pc) scales \citep{Mason06,Rhee06}. Beyond that work, the spectral properties of the small-scale extended MIR emission have been explored in only a handful of Seyfert galaxies \citep{Roche06,Roche07,Young07}. A detailed dissection of the nuclear regions is necessary to establish how --- if at all --- these results fit within AGN unified schemes.

At a distance of 32 Mpc ($H_{0}=72$ km sec$^{-1}$ Mpc$^{-1}$; 1\arcsec=150 pc), NGC~2110 is the nearest known type 2 AGN  \citep{Bradt78,McClintock79} with silicate emission \citep{Shi06}, and is unique in being both bright enough for ground-based mid-IR spectroscopy, and close enough for spatial resolution of a few $\times$ 10 parsecs to be attainable. To investigate the nature and location of the region(s) giving rise to silicate emission in active galaxies, we present ground-based and Spitzer mid-IR imaging and spectroscopy of the nucleus of NGC~2110.

\section{Observations}
\label{sec:obs}

N band spectroscopy and an imaging observation in the 11.2~$\mu$m N$^{\prime}$ filter ($\Delta \lambda = \; 2.4 \; \mu$m, 50\% cut-on/off) were acquired using Michelle \citep{Glasse97}, the mid-IR imager/spectrograph on the Gemini North telescope, on 20070318 and 20070320\footnote{Program ID: GN-2007A-Q-49}. Conditions were clear and dry (PWV $<$ 1~mm) on both nights and a standard chop-nod observing scheme was employed for both imaging and spectroscopy. The 2-pixel (0.36\arcsec) slit was used, giving spectral resolution $\lambda /  \Delta \lambda \sim 200$ and spectral coverage of the whole N band atmospheric window. The slit was oriented at 160\deg, approximately along the inner ionization cones of the galaxy \citep[][]{Pogge89,Mulchaey94}. Because of an instrumental problem at the time of the observation, the galaxy was not optimally centered in the slit. Nonetheless, an adequate signal-to-noise ratio was achieved and the pointlike nature of the nucleus (below) means that the conclusions of this work are not affected. On-source exposure times were 1200 sec for the spectroscopy and 150 sec for the imaging.

The galaxy appears pointlike in the short imaging observation, with FWHM$\approx$0\arcsec.42 (comparable to the FWHM of the photometric standard star) and radial profile well fit with a Moffat profile. There is no evidence of significant low-level extended emission in the smoothed data, implying that the Michelle spectrum represents the unresolved nuclear source.
The total flux density of the nuclear point source
was measured to be 286 mJy; variable atmospheric transmission in the mid-IR means that this value is likely accurate to $\sim$10\%. The Michelle photometry agrees with the IRS spectrum (Fig.~\ref{fig:specs1}; Shi et al. 2006; Gorjian et al. 2009, in prep) to within the uncertainties \footnote{IRS spectroscopic flux calibration is accurate to $\sim$ 5-10\%; IRS Data Handbook, version 3.1}.

Initial IRAF processing of the Michelle spectra involved combination of the chop- and nod-subtracted data, and use of a median subtraction algorithm to remove channel-channel offsets and low-level vertical striping. The spectrum was optimally extracted using the Starlink Figaro package and wavelength calibrated using sky emission lines in the raw frames. The galaxy spectrum was then divided by the spectrum of a G8III star (HR1784) to cancel telluric absorption lines, multiplied by a 4960~K blackbody curve, and flux calibrated using the imaging data. 

\section{Results and discussion}
\label{sec:results}

\begin{figure}[t]
\includegraphics[scale=0.65,angle=270]{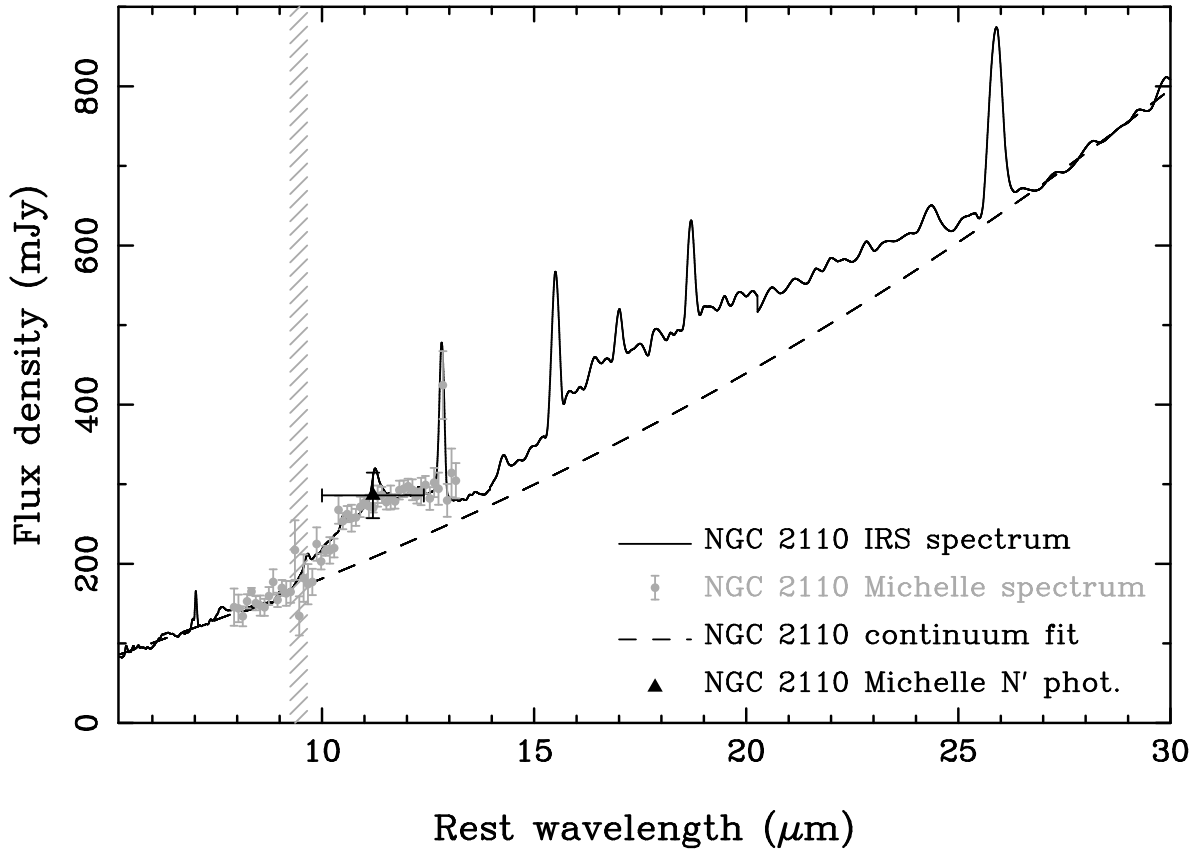}
\caption{Spitzer/IRS and Gemini/Michelle spectra of NGC~2110 (3.7-10.7\arcsec\ and 0.36\arcsec\ slits, respectively). The hatched area shows the telluric O$_{3}$ band, which is not well removed by division by the standard star. The Michelle spectrum was calibrated using the N$^{\prime}$ imaging but has been scaled slightly for ease of comparison with the IRS data.}
\label{fig:specs1}
\end{figure}

The Michelle and IRS data on NGC~2110 are presented in Fig.~\ref{fig:specs1}, with continuum-subtracted spectra in Fig.~\ref{fig:specs2}. The continuum shown in Fig.~\ref{fig:specs1} is a spline fit to the IRS spectrum  at 5 - 9~$\mu$m and 24 - 30 $\mu$m. Although fitting only to the extremes of the spectrum may exaggerate the strength of the silicate emission \citep{Sirocky08}, we use this fit for ease of comparison with previous work. The same continuum was also subtracted from the Michelle spectrum after a slight scaling to match the flux density in that spectrum.

Outside the 11.3~$\mu$m PAH band, the 8-13~$\mu$m spectral region changes very little between the Michelle and IRS spectra; the flux densities of the continuum and silicate emission in the Michelle and IRS spectra agree to within the uncertainties (\S\ref{sec:obs}}). The similarity of the strength and profile of the 10~$\mu$m silicate feature in both spectra implies that the silicate emission in NGC~2110 arises within the point source detected in the ground-based imaging, whose FWHM $<$ 63 pc.  At that distance the emission could arise in the torus itself, whose outer radius is thought to be no more than a few pc \citep{Jaffe,Packham05,Tristram07}, or in the inner part of the narrow line region, which extends to $r\sim230$ pc in the optical \citep{Mulchaey94}. The compactness of the emission argues against an origin in mass-losing evolved stars, as detected in Virgo cluster galaxies \citep{Bressan06}.

Fig.~\ref{fig:specs2} also shows the silicate features in the LINER/type 1 Seyfert galaxy, NGC~3998 \citep{Sturm05}. The features in NGC~3998 are weaker than those in NGC~2110 but the profiles and peak wavelengths of their 10~$\mu$m bands are quite similar. However, the 18~$\mu$m feature in NGC~2110 is weaker relative to the 10~$\mu$m band than in NGC~3998, and NGC~2110 exhibits a more pronounced red wing on the 18~$\mu$m feature. In this respect, NGC~2110 bears more resemblance to the QSOs in \citet{Hao05}, which also have strong red wings to their 18~$\mu$m bands. The peak height of the 18~$\mu$m feature is sensitive to the choice of continuum, but other reasonable continuum fits give similar 10~$\mu$m profiles and also show the red wing on the 18~$\mu$m band. The properties (profiles, peak wavelengths and relative strengths) of the silicate bands in NGC~2110 lie within the range exhibited by other AGN with silicate emission features.
\citet{Sturm05} note that the 10~$\mu$m band in NGC~3998 is broadened and shifted to longer wavelengths than in the Galactic ISM, and discuss grain size and composition effects that could explain the observed spectrum. The similarity of the 10~$\mu$m band in NGC~2110 to that in NGC~3998 suggests that such grain processing may also be at work in this Seyfert 2 nucleus.

\citet{Schweitzer08} have modeled silicate emission arising from dust in the NLR of PG QSOs. The distance to the dust giving rise to the feature depends on various model parameters (e.g. NLR density, extinction),  but on average  $R_{\rm dust} \approx 80(L_{\rm bol46})^{1/2} \rm pc$, where $L_{\rm bol} = 7L(5100)$ and is in units of $10^{46} \rm erg \; s^{-1}$. For the silicate-emitting dust to be $>$ 32 pc from the nucleus would require $L_{\rm bol} \gtrsim 1.4 \times 10^{45} \rm erg \; sec^{-1}$.  No 5100$\AA$ continuum luminosity is available for NGC~2110, but \citet{Moran07} find $L_{\rm bol} \approx 2 \times 10^{44} \rm erg \; sec^{-1}$ based on the extinction-corrected [OIII] $\lambda$5007 line. This implies  $R_{\rm dust} < 11 \rm \; pc$, consistent with the constraint on the distance to the silicate emitting dust imposed by the comparison of the Michelle and IRS data. This distance is clearly much smaller than the optical size of the NLR in NGC~2110, an observation also noted for their PG QSO sample by Schweitzer et al., who suggest that the silicate-emitting dust may be associated with highly ionised gas in the innermost regions of the NLR. 

The calculations of \citet{Schweitzer08} demonstrate that the silicate
emission in AGN may arise in the NLR, but they do not rule out that
some or all of it could be produced by dust in the torus (and there
need not in fact be a physical difference between the outer edge of a
clumpy torus and the beginning of the NLR). For clumpy tori,
\citet{Nenkova08b} find that if the optical depth per cloud
$\tau_{V}>100$ or if the number of clouds along the line of sight is
$N_{0}\sim 2$, a silicate feature can be produced even for lines of
sight close to the equatorial plane.

We fit the  clumpy torus models of
\citet{Nenkova08a} to the MIR spectra of NGC~2110.  In these radiative transfer computations, the
individual clouds have a fixed optical depth.  The additional
parameters detail the distribution of the clouds.  They are radially
distributed according to a power law, $\propto r^{-q}$, from the dust
sublimation radius, $R_d$ to an outer radius $R_o$, which we
parameterize with $Y={R_{o}}/{R_{d}}$.  The torus has width parameter
$\sigma$ with a Gaussian edge.  The average number of clouds along an
equatorial ray is $N_0$. The number along a given line of sight
declines with altitude, with $N_{los}(\beta) = N_0
\exp(-\beta^{2}/\sigma^{2})$ along angle $\beta$ from the equator,
and the emergent reprocessed emission is calculated for all viewing
angles, $i$.

\begin{figure}
\centerline{\includegraphics[width=6in]{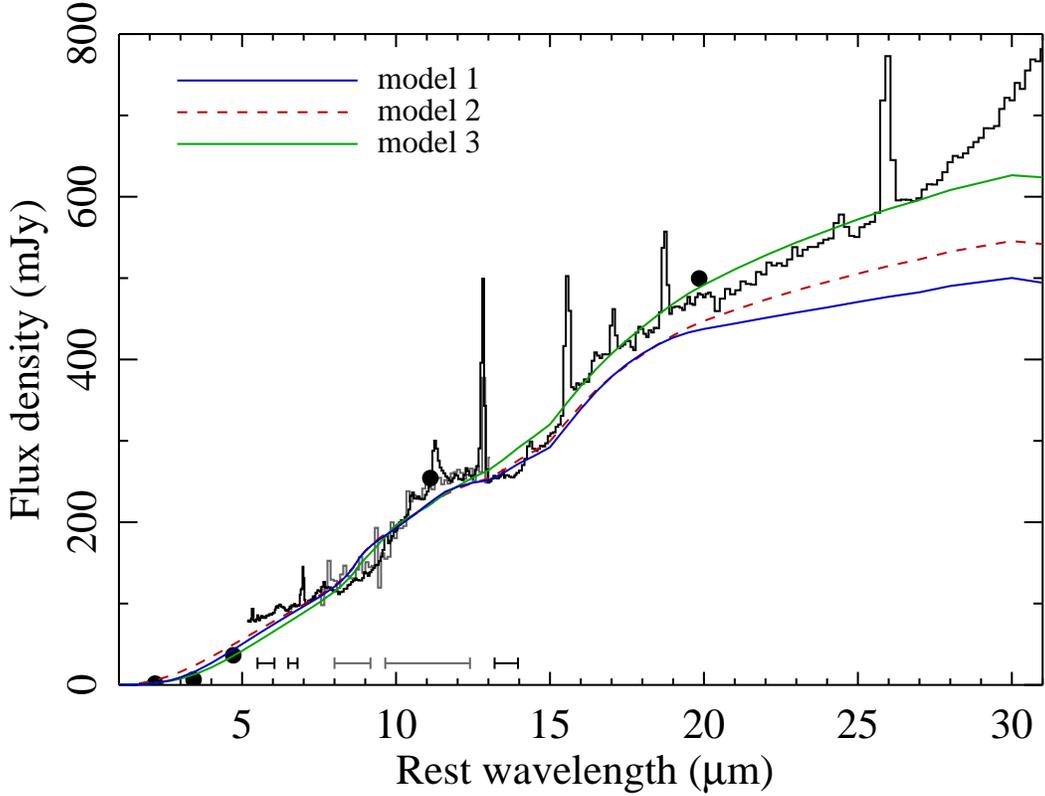}}
\caption{\label{fig:modbest}
Clumpy torus models applied to the MIR spectra.
Bars mark the regions
employed in the fitting; IRS SL data at the short and
long wavelength ends (black), and Michelle
measurements from 8--12.5$\mu$m (gray).
Photometric measurements \citep[][this work]{Lawrence85,AlonsoHerrero96} and the IRS LL spectrum are plotted but not used in the fitting.
The parameters of the best-fitting model 1 and
model 2 are similar, with a thin torus ($\sigma = 15^\circ$),
and limited optical depth through individual clouds ($\tau_V = 20$),
but the BLR remains obscured. 
Model 3 has a  larger optical depth per cloud,
 a larger width parameter ($\sigma = 45^\circ$), and fewer clouds along
the torus equator.
}
\end{figure}

We fit only the MIR continuum, using the relatively line-free Michelle
spectrum from 8 to 12.5~$\mu$m (except the ozone region) and the 
IRS spectrum in regions indicated in Fig.~\ref{fig:modbest}.
Because of uncertainties in starlight subtraction
\citep{AlonsoHerrero96} and the possible variability of this object in
the NIR \citep{Lawrence85}, we do not fit the photometric data
points. Neither do we fit to the LL spectrum of NGC~2110; although most of the 11.2~$\mu$m
flux arises in a source with FWHM$<$0.42\arcsec, contributions from
extended emission may become significant at longer wavelengths.

We allow all the model parameters to be free.  One disadvantage of the
models is that they are highly degenerate.  Concentrating
on the MIR, where only the clouds close to the AGN (within $\sim 15
R_d$) are relevant, a number of different models produce similar fits
and spectral shapes.  The total cloud distribution determines the MIR
emission, so various combinations of total number of clouds, outer
extent, and radial profile produce similar spectra.  The best-fitting
models all have near-equatorial views through the torus, which
has a small width parameter. The inclination of the torus is driven by
the NIR/MIR ratio in the data; more face-on models cannot
fit the short end of the IRS spectrum without predicting less
14~$\mu$m emission than is observed. Larger tori may produce a slightly lower NIR/MIR ratio \citep{Nenkova08b} but large tori are inconsistent with high resolution observations \citep[][although see Kishimoto et al. 2009 for contrary evidence]{Jaffe,Tristram07}. 

We plot the formally best-fitting model (``model 1'') in Figure
\ref{fig:modbest}, which has $\tau_V = 20$, $Y = 30$, $q = 0$, $N_0 =  
5$, $\sigma = 15^\circ$, and $i = 80^\circ$.  A similar model (``model 
2") changes $N_o$ to 10 and $i$ to $ 70^\circ$.  The tori have smaller opening angles than required by the unified model
to explain the relative numbers of types 1/2 AGN. However, slender
tori are suggested by the disk-like broad line region (BLR) which is
detected in polarized light \citep{Moran07}. The apparent scarcity of
type 2 AGN with silicate emission may also indicate that these objects
are outliers in terms of torus properties.  A somewhat different
solution (``model 3'') has fewer clouds along a given line of sight,
and the torus covers a larger angular extent ($\tau_V = 40$, $Y = 30$,
$q = 0$, $N_0 =2$, $\sigma = 45^\circ$, and $i = 90^\circ$).  All
these models obscure the central engine, with $N_H \approx 1$--$2
\times 10^{23} \mathrm{\, cm^{-2}}$ along the line of sight, assuming
a standard gas-to-dust ratio. This $N_H$ is within a factor of a few of that estimated by most authors \citep{Malaguti99,Risaliti02,Evans07}

The vertical scaling is a free parameter and sets the bolometric
luminosity of the model source.  We find $L_{bol} = 4$, 3, and 0.7
$\times 10^{44} \mathrm{\, erg \, s^{-1}}$ in models 1, 2, and 3, respectively.
 These values are roughly
consistent with the $L_{\rm bol} \approx 2 \times 10^{44} \rm erg \; sec^{-1}$ estimated from [OIII] $\lambda$5007 \citep{Moran07}.
In all cases, the silicate-emitting region is small.
For a standard AGN heating spectrum and a dust sublimation temperature
of 1500 K,
$R_d = 0.4 (L_{bol}/10^{45})^{1/2}$ pc.
Thus, the outer extent of the torus we model, at $30\; R_d$,
is $<$ 8 pc.

\begin{figure}
\centerline{\includegraphics[width=6in]{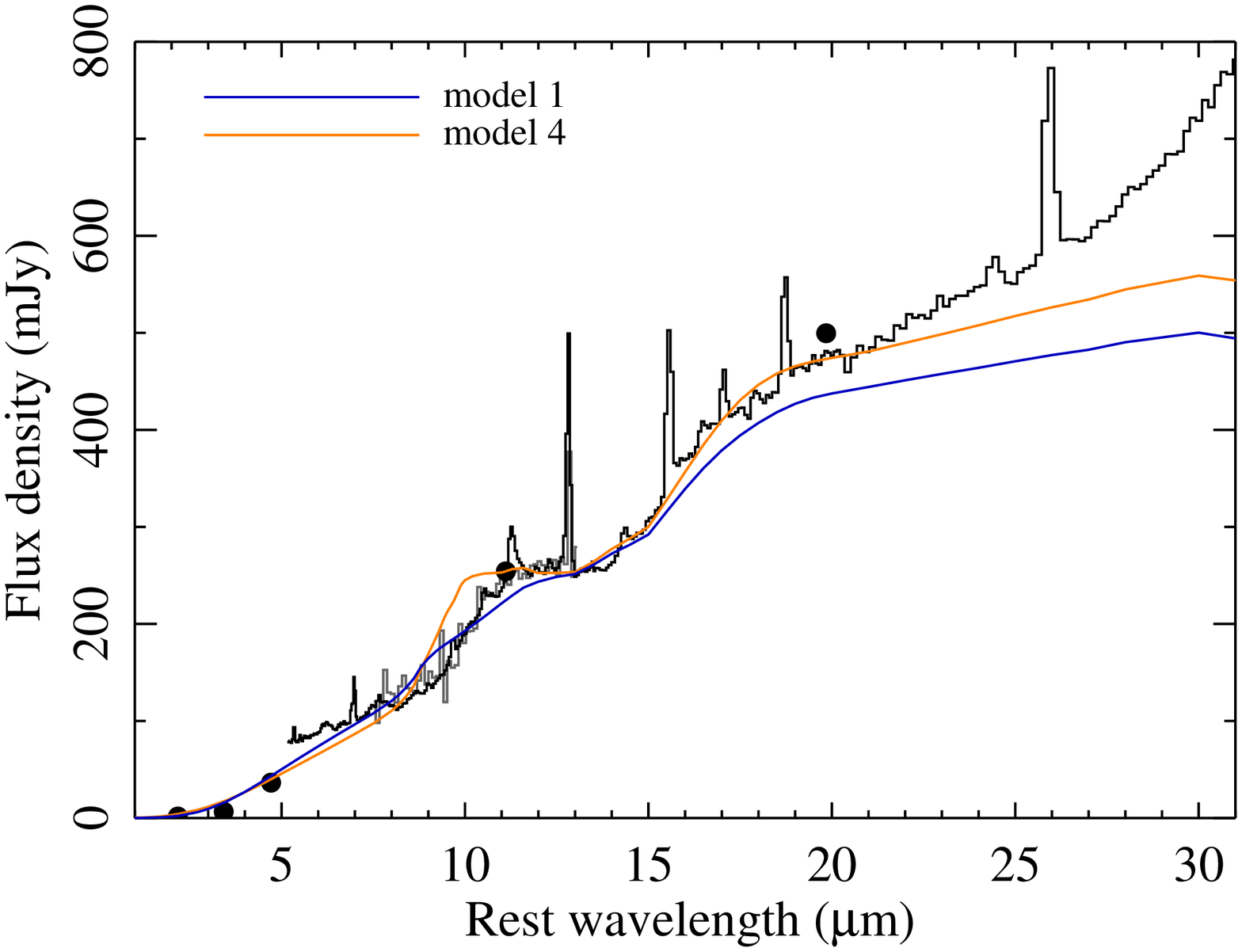}}
\caption{\label{fig:modalt}
The clumpy torus can produce strong silicate emission even
when the direct view of the AGN is blocked (model 4; $i = 80^{\circ}$).
This model yields
 a formally poor fit because the
wavelength of the silicate features depends on the
dust composition and grain size, which we do not vary.
}
\end{figure}

One problem with all these models is that they fail to capture the
steep rise of the silicate feature longward of 10~$\mu$m. The 
wavelength of the silicate peak is a function of the dust composition.
All the models we present use the cold astronomical silicate of \citet{Ossenkopf92}, which has a maximum optical depth at
10.0~$\mu$m.  We do not experiment with alternate dust compositions,
but we do demonstrate in Figure \ref{fig:modalt} that even an obscured AGN may exhibit strong
silicate emission.  ``Model 4'' is
similar to models 1 and 2, with a slender torus ($\sigma =15^\circ$).
The primary difference is that it contains fewer clouds along radial
rays, but with nearly equatorial viewing ($i = 80^{\circ}$) and
$\tau_V = 40$, the average line-of-sight optical depth is still $1
\times 10^{23} \mathrm{\, cm^{-2}}$.
Although this model is formally a poor fit, it has the advantage that the shape of the 10~$\mu$m silicate feature does not show the ``double-peaked,'' self-absorbed structure seen in models 1--3. 
The dust composition determines the shape and strength of the silicate feature, while variations of the grain size affect the relative extinction of the continuum.  A distribution favoring large grains does not alter these models significantly.  Favoring small grains corresponds to smaller values of $\tau_{\rm V}$ for the same silicate strength and profile.  The NIR extinction is then reduced relative to the MIR, so the preference for edge-on views remains
robust.

To summarize, these
observations permit us for the first time to set tight limits (r$<$32 pc) on the size of the silicate-emitting region in a Seyfert galaxy. This indicates that any silicate emission from the
NLR must arise in its innermost regions. Alternatively, the silicate emission could come from the torus:  we show that clumpy torus models give a reasonable fit to the silicate emission and the 2-20 $\mu$m SED, while at the same time obscuring the BLR. We emphasize that in the context of a clumpy torus
the distinction of the transition from the outer torus to the inner NLR is more semantic than physical.  Simultaneous modeling of emission lines and silicate emission from the NLR, as proposed by \citet{Schweitzer08}, may further illuminate the origin of the silicate emission features. Measurements on small spatial scales remain essential to identify the emitting structures in the cores of active galaxies.

\acknowledgments

We thank E.~Sturm for providing the spectrum of NGC~3998 and the referee, S.~H\"{o}nig, for a timely and helpful report. Based on observations obtained at the Gemini Observatory, which is operated by the
Association of Universities for Research in Astronomy, Inc., under a cooperative agreement
with the NSF on behalf of the Gemini partnership: the National Science Foundation (United
States), the Science and Technology Facilities Council (United Kingdom), the
National Research Council (Canada), CONICYT (Chile), the Australian Research Council
(Australia), MinistŽrio da Cincia e Tecnologia (Brazil) and SECYT (Argentina). This work is based in part on observations made with the Spitzer Space Telescope, which is operated by the Jet Propulsion Laboratory, California Institute of Technology under a contract with NASA.
N.A.L. acknowledges work supported by the NSF under Grant 0237291 and the hospitality of 
the University of Florida Department of Astronomy.

%%
%% thebibliography produces citations in the text using \bibitem-\cite
%% cross-referencing. Each reference is preceded by a
%% \bibitem command that defines in curly braces the KEY that corresponds
%% to the KEY in the \cite commands (see the first section above).
%% Make sure that you provide a unique KEY for every \bibitem or else the
%% paper will not LaTeX. The square brackets should contain
%% the citation text that LaTeX will insert in
%% place of the \cite commands.

%% We have used macros to produce journal name abbreviations.
%% AASTeX provides a number of these for the more frequently-cited journals.
%% See the Author Guide for a list of them.

%% Note that the style of the \bibitem labels (in []) is slightly
%% different from previous examples.  The natbib system solves a host
%% of citation expression problems, but it is necessary to clearly
%% delimit the year from the author name used in the citation.
%% See the natbib documentation for more details and options.

%\bibliographystyle{/Users/rmason/aastex502/apj}
%\bibliography{/Users/rmason/aastex502/refs}

%Contents of .bbl pasted in for use with ApJL length calculator, also for submission if refs don't change.

\end{document}